\documentstyle[aps,multicol,epsf]{revtex}

\def\sech{\,\hbox{sech}}

\begin{document}
\title{
  \begin{flushright} \begin{small}
    hep-th/0005236
  \end{small} \end{flushright}
\vspace{1.cm}
Anisotropic Four-Dimensional NS-NS String Cosmology}
\author{Chiang-Mei Chen\footnote{E-mail: cmchen@joule.phy.ncu.edu.tw}}
\address{Department of Physics, National Central University,
         Chungli 320, Taiwan}
\author{T. Harko\footnote{E-mail: tcharko@hkusua.hku.hk}}
\address{Department of Physics, University of Hong Kong,
         Pokfulam, Hong Kong}
\author{M. K. Mak\footnote{E-mail: mkmak@vtc.edu.hk}}
\address{Department of Physics, Hong Kong University of Science
         and Technology, Clear Water Bay, Hong Kong}
\date{May 25, 2000}
\maketitle

\begin{abstract}
An anisotropic (Bianchi type I) cosmology is considered in the
four-dimensional NS-NS sector of low-energy effective string theory
coupled to a dilaton and an axion-like $H$-field within a de
Sitter-Einstein frame background.  The time evolution of this Universe
is discussed in both the Einstein and string frames.
\end{abstract}

\vspace{.5cm}
\hspace{1cm}{PACS number(s): 04.20.Jb, 04.65.+e, 98.80.-k}

\begin{multicols}{2}
\narrowtext

Pre-Big Bang cosmological models \cite{GaVe93}, based on the low
energy limit of the string theory, have been intensively investigated
in the recent physics literature \cite{DiDe99}-\cite{DuSa00}.
Generically, in these type of models the dynamics of the Universe is
dominated by massless bosonic fields.  In the string frame, the
four-dimensional NS-NS effective action, which is common to both the
heterotic and type II string theories, is given by
\cite{CaMaPeFr85}-\cite{Ro86}
\begin{equation} \label{SS}
\hat {\cal S} = \int d^4x \sqrt{-\hat g} e^{-2\phi} \left\{ \hat R
   + \hat \kappa (\nabla \phi)^2 - \frac1{12} \hat H_{[3]}^2
   - \hat U \right\},
\end{equation}
where $H_{\mu\nu\lambda}=\partial_{[\mu}B_{\nu\lambda]}$ is an
antisymmetric tensor field, $\hat \kappa$ is a generalized dilaton
coupling constant and $\hat U=\hat U(\phi)$ is a dilaton potential.
In addition, $\hat H_{[3]}^2$ means the square of the $H$-field with
respect to the metric $\hat g_{\mu\nu}$.  The low-energy string action
posseses a symmetry property, called scale factor duality, which lets
us expect that the present phase of the Universe is preceded by an
inflationary pre-Big Bang phase.  Explicit dual solutions can be
constructed for each Bianchi space-time, except the Bianchi class A
types VIII and IX models \cite{DiDe99}.

By means of the conformal rescaling
\begin{equation}\label{g2g}
g_{\mu\nu} = e^{-2\phi} \, \hat g_{\mu\nu},
\end{equation}
the action (\ref{SS}) can be transformed to the so-called Einstein
frame as
\begin{equation} \label{SE}
{\cal S} = \int d^4x \sqrt{-g} \left\{ R - \kappa (\nabla \phi)^2
   - \frac1{12} e^{-4\phi} \, H_{[3]}^2 - U \right\},
\end{equation}
where $\kappa=6-\hat \kappa$, $U=e^{2\phi}\,\hat U$ and
$H_{[3]}^2$ denotes the square of the antisymmetric field by
$g_{\mu\nu}$.

Analytic biaxial (two scale factors only) Bianchi type I geometry has
been previously considered in \cite{CoLaWa94} for the case with
nonvanishing $H_{[3]}$ but without a dilaton field potential, i.e.
$U \equiv 0$.  Triaxial models with the central deficit charge
constrained to zero in the presence of a modulus field (representing
the evolution of compact extra dimensions) have been analyzed in
\cite{CoLaWa95}.  Recently, a study of spatially flat and homogeneous
string cosmologies, considering the combined effects of the dilaton,
modulus, two-form potential and central charge deficit, and using
methods from the qualitative theory of differential equations (phase
portrait analysis) has been presented in \cite{BiCoLi99},

The general Bianchi type I space-time for arbitrary dimensional
dilaton gravities, with vanishing antisymmetric tensor
$H_{\mu\nu\lambda}$ and in the presence of an exponential type
dilaton field potential, have been obtained in both the Einstein and
string frames \cite{ChHaMa00}.

It is the purpose of the present letter to consider, in the framework
of a four-dimensional Bianchi type I geometry, the effects on the
dynamics and evolution of the early Universe of a non-vanishing
antisymmetric field and of a string frame exponential type dilaton
field potential.

In the Einstein frame the field equations, which follow from variation
of (\ref{SE}), are given by
\begin{eqnarray}
R_{\mu\nu} - \kappa \partial_\mu \phi \partial_\nu \phi
   - \frac12 g_{\mu\nu} U \qquad\qquad && \nonumber\\
   - \frac14 e^{-4\phi} \left( H_{\mu\alpha\beta} H_\nu{}^{\alpha\beta}
   - \frac13 g_{\mu\nu} H^2 \right) &=& 0, \label{EqR} \\
\nabla_\mu \left( e^{-4\phi} H^{\mu\nu\lambda} \right) &=& 0, \label{EqH}\\
\nabla^2 \phi + \frac1{6\kappa} e^{-4\phi} H^2
   - \frac1{2\kappa} \frac{\partial U}{\partial \phi} &=& 0. \label{Ephi}
\end{eqnarray}
Moreover, the $H$-field must satisfy the integrability condition
(Bianchi identity) $\partial_{[\mu} H_{\nu\lambda\rho]}=0$.

In four dimensions, every three-form field can be dualized to a
pseudoscalar.
Thus, an appropriate ansatz for the $H$-field is \cite{CoLaWa94}
\begin{equation}\label{Deh}
H^{\mu\nu\lambda} = \frac1{\sqrt{-g}} \, e^{4\phi} \,
    \epsilon^{\mu\nu\lambda\rho} \, \partial_\rho h,
\end{equation}
where $\epsilon^{\mu\nu\lambda\rho}
=-\delta^\mu_{[0} \delta^\nu_1 \delta^\lambda_2 \delta^\rho_{3]}$
is the total antisymmetric tensor and $h=h(t)$ is the Kalb-Ramond
axion field.  Then the field equation (\ref{EqH}) is satisfied
automatically and the Bianchi identity becomes
\begin{equation}\label{Eh}
\partial_{\mu} \left( \sqrt{-g} \, e^{4\phi} \, \partial^\mu h \right) = 0.
\end{equation}
Moreover, we shall assume that in the string frame the dilaton field
potential is of exponential type
\begin{equation}\label{AU}
\hat U(\phi) = \Lambda \, e^{-2\phi},
\end{equation}
with $\Lambda$ a non-negative constant (de Sitter space-time).
Therefore in the Einstein frame the effect of the potential is identical
to that of a cosmological constant, $U(\phi)=\Lambda$.

For the Bianchi type I space-time, in the Einstein frame,
\begin{equation}
ds^2 = -dt^2 + \sum_{i=1}^3 a_i^2(t) \left( dx^i \right)^2,
\end{equation}
and the ansatz (\ref{Deh},\ref{AU}), the field equations
(\ref{EqR},\ref{Ephi},\ref{Eh}) take the form
\begin{eqnarray}
3 \dot \theta + \sum_{i=1}^3 \theta_i^2 + \kappa \dot \phi^2
   + \frac12 e^{4\phi} \, \dot h^2 - \frac12 \Lambda &=& 0, \label{dth} \\
\frac1{V} \frac{d}{dt} ( V \theta_i ) - \frac12 \Lambda &=& 0, \,\,
     i=1,2,3, \label{dV} \\
\ddot h + 3 \theta \dot h + 4 \dot \phi \dot h &=& 0, \label{dh} \\
\frac1{V} \frac{d}{dt} ( V \dot\phi )
   - \frac1{\kappa} e^{4\phi} \dot h^2 &=& 0,
\end{eqnarray}
where we have introduced the volume scale factor,
$V := \prod_{i=1}^3 a_i$,
directional Hubble factors,
$\theta_i := \dot a_i / a_i, \, i=1,2,3$,
and the mean Hubble factor,
$\theta := \sum_{i=1}^3 \theta_i/3 = \dot V/3V$.
We shall also introduce two basic physical observational quantities in
cosmology: the mean anisotropy parameter,
$A := \sum_{i=1}^3 (\theta_i-\theta)^2/3\theta^2$,
and the deceleration parameter, $q=\frac{d}{dt} \theta^{-1}-1$.

By summing equations (\ref{dV}) we obtain
\begin{equation}\label{SV}
\frac1{V} \frac{d}{dt} ( V \theta ) = \frac12 \Lambda,
\end{equation}
which, together with (\ref{dV}), leads to
\begin{equation}
\theta_i = \theta + K_i V^{-1}, \quad i=1,2,3,
\end{equation}
with $K_i, \, i=1,2,3$ being constants of integration satisfying
$\sum_{i=1}^3 K_i = 0$.

It is worth noticing that, in this framework, the geometry of the
considered Universe, which is described by $a_i(t),\,i=1,2,3$, is
determined only by the existence of the cosmological constant
$\Lambda$ and is ``decoupled'' from the matter fields $\phi$ and $H$.
(The effect of matter fields is presented in the magnitude of the
parameters, i.e. constants of integration.)

From equation (\ref{SV}) we obtain the time evolution of the
mean Hubble factor,
\begin{equation}
\theta(\tau) = \sqrt{\frac{\Lambda}6} \coth \tau,
\end{equation}
leading to
\begin{eqnarray}
V(\tau) &=& V_0 \sinh \tau, \\
a_i(\tau) &=& a_{i0} \sinh^{\alpha_i^+} \frac{\tau}2
     \cosh^{\alpha_i^-} \frac{\tau}2, \quad i=1,2,3,
\end{eqnarray}
where $\tau:=\sqrt{3\Lambda/2}(t-t_0)$ and
$\alpha_i^\pm := 1/3 \pm \sqrt{2/3\Lambda}K_i/V_0$.
The mean anisotropy and the deceleration parameter are given by
\begin{eqnarray}
A(\tau) &=& \frac{2K^2}{\Lambda V_0^2} \sech^2 \tau, \\
q(\tau) &=& 3 \sech^2 \tau - 1,
\end{eqnarray}
where $K^2=\sum_{i=1}^3 K_i^2$.

Equation (\ref{dh}) can be integrated to give
\begin{equation}
\dot h = C \, e^{-4\phi} \, V^{-1},
\end{equation}
with $C$ a constant of integration.
Thus the dynamics of the dilaton field in the Einstein frame is
described by the following differential equation
\begin{equation}
\sinh \tau \frac{d}{dt} \left( \sinh\tau\,\dot\phi \right)
   = \frac{C^2}{\kappa V_0^2} e^{-4\phi},
\end{equation}
with the general solution
\begin{equation}
e^{2\phi(\tau)} = \varphi_0^2 \left( \tanh^\omega\frac{\tau}2
   + \tanh^{-\omega}\frac{\tau}2 \right),
\end{equation}
where we denote $\omega:=\sqrt{8\phi_0/3\Lambda}/V_0$,
$\varphi_0^2:=\sqrt{C^2/8\kappa\phi_0}$
and $\phi_0>0$ is a constant of integration.
The antisymmetric tensor field is given by
\begin{equation}
h(\tau) = h_0 + \frac{\kappa\sqrt{\phi_0}}{C} \,
    \frac{\tanh^{2\omega}\frac{\tau}2-1}
         {\tanh^{2\omega}\frac{\tau}2+1},
\end{equation}
with $h_0$ an arbitrary constant.

The integration constants must satisfy the consistency condition,
\begin{equation}
K^2 = \Lambda V_0^2 - \kappa \phi_0,
\end{equation}
which follows from equation (\ref{dth}).

In the case of vanishing cosmological constant, $\Lambda \equiv 0$,
the general solution in the Einstein frame of the gravitational field
equations for a Bianchi type I geometry with dilaton and Kalb-Ramond
axion fields is given by:

\begin{eqnarray}
\theta(t) &=& \frac1{3t}, \qquad V(t) = V_0 t, \\
a_i(t) &=& a_{i0} \, t^{1/3+K_i/V_0}, \quad i=1,2,3, \\
A &=& 3 K^2 V_0^{-2} = const., \qquad q = 2 = const., \\
e^{2\phi(t)} &=& \varphi_0^2 \,
    \left( t^\alpha + t^{-\alpha} \right), \\
h(t) &=& h_0 - \frac{2\kappa\sqrt{\phi_0}}{C}
    \left( t^{2\alpha} + 1 \right)^{-1},
\end{eqnarray}
together with the consistency condition
\begin{equation}
K^2 = \frac23 V_0^2 - \kappa \phi_0,
\end{equation}
where $\alpha=2\sqrt{\phi_0}/V_0$.

In order to find the general solution of the gravitational field
equations in the string frame with the line element
\begin{equation}
d\hat s^2 = - d\hat t^2
   + \sum_{i=1}^3 \hat a_i^2(\hat t) \left( dx^i \right)^2,
\end{equation}
we must perform the conformal transformation (\ref{g2g}).  To obtain a
simpler mathematical form of the equations we shall introduce a new
variable $\eta=\tanh\tau/2$, $\eta \in [0,1]$ and denote $\hat
\varphi_0:= \sqrt{8/3\Lambda}\,\varphi_0$.  Then the string
frame time evolution of the Bianchi type I space-time with dilaton and
Kalb-Ramond axion fields and an exponential type dilaton potential can
be expressed in the following exact parametric form:
\begin{eqnarray}
\hat t(\eta) &=& \hat t_0 + \hat \varphi_0 \int
    \frac{\sqrt{\eta^\omega+\eta^{-\omega}}}{1-\eta^2} \, d\eta, \\
\hat V(\eta) &=& \hat V_0 \, ( \eta^\omega+\eta^{-\omega} )^{3/2}
    \frac{\eta}{1-\eta^2}, \\
\hat \theta(\eta) &=& \frac1{3\hat \varphi_0}
    \frac{1-\eta^2}{\eta \sqrt{\eta^\omega\!+\!\eta^{-\omega}}}
    \left( \frac{3\omega}2
      \frac{\eta^\omega\!-\!\eta^{-\omega}}{\eta^\omega\!+\!\eta^{-\omega}}
      \!+\! \frac{1\!+\!\eta^2}{1\!-\!\eta^2} \right), \\
\hat a_i(\eta) &=& \hat a_{i0}
    \frac{\eta^{\alpha_i^+}\sqrt{\eta^\omega+\eta^{-\omega}}}
         {(1-\eta^2)^{1/3}} \quad i=1,2,3, \\
\hat A(\eta) &=& \frac{2K^2}{\Lambda V_0^2} \left(
    \frac{3\omega}2 \,
          \frac{\eta^\omega-\eta^{-\omega}}{\eta^\omega+\eta^{-\omega}}
        + \frac{1+\eta^2}{1-\eta^2} \right)^{-2}, \\
\hat q(\eta) &:=& \frac{d}{d\hat t}\hat \theta^{-1} - 1
   = \frac{1-\eta^2}{\hat \varphi_0\sqrt{\eta^\omega+\eta^{-\omega}}}
     \, \frac{d}{d\eta} \hat \theta^{-1} - 1.
\end{eqnarray}

In the case of a vanishing cosmological constant the string frame
solution of the gravitational field equations with dilaton and axion
fields is given again in a parametric form by:
\begin{eqnarray}
\hat t(t) &=& \hat t_0 + \varphi_0 \,
    \int \sqrt{t^\alpha + t^{-\alpha}} \, dt, \\
\hat V(t) &=& \hat V_0 \, t \, ( t^{\alpha}+t^{-\alpha})^{3/2}, \\
\hat \theta(t) &=& \varphi_0^{-1} \, \left(
    \frac{\alpha}2 \, \frac{t^\alpha-t^{-\alpha}}{t^\alpha+t^{-\alpha}}
   + \frac13 \right) \frac1{t(t^\alpha+t^{-\alpha})^{1/2}}, \\
\hat a_i(t) &=& \hat a_{i0} \, t^{1/3+K_i/V_0}
    \sqrt{t^\alpha+t^{-\alpha}}, \quad i=1,2,3, \\
\hat A(t) &=& \frac{K^2}{3V_0^2} \, \left(
    \frac{\alpha}2 \frac{t^\alpha-t^{-\alpha}}{t^\alpha+t^{-\alpha}}
   + \frac13 \right)^{-2}, \\
\hat q(t) &=& 2 - \frac{\alpha \left( \frac13 \,
            \frac{t^\alpha-t^{-\alpha}}{t^\alpha+t^{-\alpha}}
          + \frac{\alpha}2 \right)}{\left( \frac{\alpha}2 \,
            \frac{t^\alpha-t^{-\alpha}}{t^\alpha+t^{-\alpha}}
          + \frac13 \right)^2}.
\end{eqnarray}

In the present letter we have presented the exact solution of the
gravitational field equations for a Bianchi type I space-time with
dilaton and axion fields in both the Einstein and string frames.  In
the Einstein frame the evolution of the Bianchi type I Universe in the
presence of a cosmological constant starts from a singular state, but
with finite values of the mean anisotropy and deceleration parameter.
In the large time limit the mean anisotropy tends to zero, $A \to 0$,
and the Universe will end in an isotropic inflationary de Sitter phase
with a negative deceleration parameter, $q<0$.  In the large time
limit the dilaton and axion fields become constants. Moreover, in the
Einstein frame, the dynamics and evolution of the Universe is
determined
only by the presence of a cosmological constant and there is no coupling
between the metric and the dilaton and axion fields.

In the string frame the dilaton and axion fields are coupled to the
metric.  Depending on the values of the constant $\omega$ there are
two distinct types of behavior.  In the first type of evolution,
corresponding to $\omega<2/3$, the Universe starts from a singular
state with zero values of the scale factors, $\hat a_i(0)=0,\,i=1,2,3$
and expands indefinitely.  In the second case, when $\omega>2/3$, the
Bianchi type I Universe starts its evolution with infinite values of
the scale factors and collapses to a bounce state, corresponding to
minimum finite non-zero values of the scale factors.  From this
non-singular state the Universe starts to expand, ending in an
isotropic and inflationary era.  The values of the physical quantities
at the bounce correspond to the values of $\eta$ satisfying the
equation $d\hat V/d\eta=0$ or
\begin{equation}
\frac{(\eta^\omega+\eta^{-\omega})^{3/2}}{1-\eta^2}
    \left( \frac{3\omega}2
      \frac{\eta^\omega-\eta^{-\omega}}{\eta^\omega+\eta^{-\omega}}
      + \frac{1+\eta^2}{1-\eta^2} \right) = 0.
\end{equation}

The string frame time variation of the volume scale factor of the
Bianchi type I space-time for different values of $\omega$ is
presented in Fig.1.  Independently of which type of evolution
classified by the value of $\omega$, in the presence of an exponential
type dilaton potential and of an axion field, the Bianchi type I
Universe always isotropizes in the large time limit, $\hat A \to 0$
for $\hat t \to \infty$.  But the dynamics of the mean anisotropy
factor is very different for the two types of evolution.  During the
collapse to the bounce the mean anisotropy increases to an infinite
value and then, during the expansionary period, tends rapidly to zero.
Hence in this case the expansionary evolution of the Bianchi type I
Universe starts with non-singular scale factors and with maximum
anisotropy.  The string frame time variation of the anisotropy
parameter and of the deceleration parameter are represented in the
Figures 2 and 3, respectively.  In the string frame and in the
presence of a dilaton potential the large time evolution is
inflationary for all times and for all $\omega$.

In the absence of a cosmological constant or a dilaton field potential
the Universe does not isotropize.  In this case the Einstein frame
mean anisotropy is constant for all times and the evolution is of the
Kasner type.  In the string frame the mean anisotropy tends, in the
large time limit, to a constant non-zero value, hence showing that the
Universe will never end in an isotropic flat Robertson-Walker type
phase.  The deceleration parameter in both frames is positive for all
times and an inflationary evolution is also impossible.  Therefore
string cosmological models involving only pure dilaton and axion
fields do not have, at least in the case of Bianchi type I anisotropic
geometries, the ability of providing realistic cosmological models.
To obtain a transition from an anisotropic state to an isotropic
inflationary one the ``good old'' cosmological constant is still the
key ingredient.

\begin{figure}
\epsfxsize=8cm
\epsffile{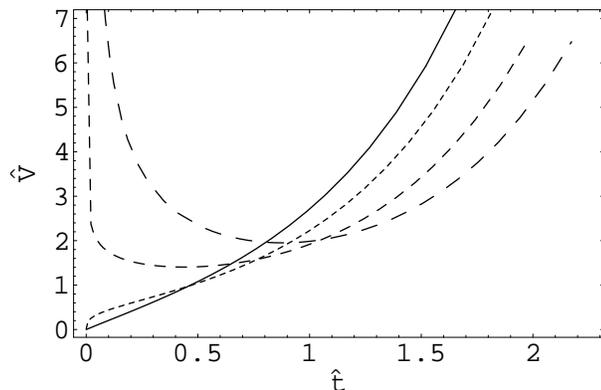}
\caption{
  String frame evolution of the volume scale factor $\hat V$ for different
  values of the parameter $\omega$:
  $\omega=0.1$ (full curve), $\omega=0.5$ (dotted curve), $\omega=0.75$
  (short dashed curve) and $\omega=0.9$ (long dashed curve).
  We have used the normalization $\hat V_0=1$ and $\hat t_0=0$}
\label{FIG1}
\end{figure}

\begin{figure}
\epsfxsize=8cm
\epsffile{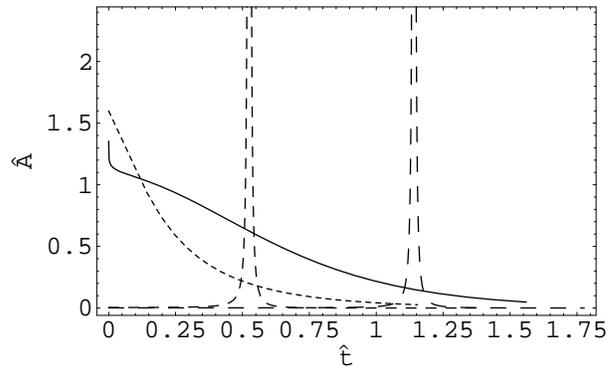}
\caption{
  String frame time variation of the mean anisotropy parameter $\hat A$
  for different values of $\omega$:
  $\omega=0.1$ (full curve), $\omega=0.5$ (dotted curve), $\omega=0.75$
  (short dashed curve) and $\omega=0.9$ (long dashed curve).
  We have used the normalization $2K^2/V_0^2 \Lambda=1$.}
\label{FIG2}
\end{figure}

\begin{figure}
\epsfxsize=8cm
\epsffile{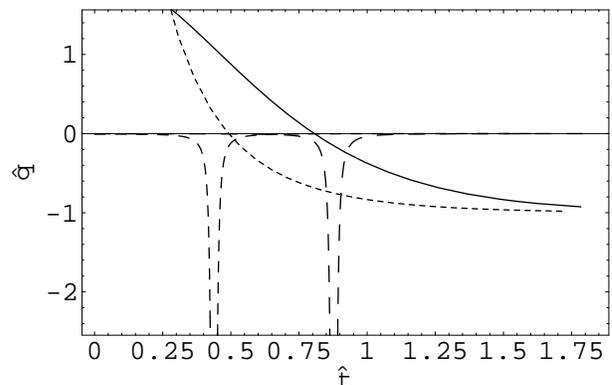}
\caption{
  Dynamics of the deceleration parameter $\hat q$ in the string frame
  for different values of $\omega$:
  $\omega=0.1$ (full curve), $\omega=0.5$ (dotted curve),
  $\omega=0.75$ (short dashed curve) and $\omega=0.9$ (long dashed curve).}
\label{FIG3}
\end{figure}

One of the authors (CMC) would like to thank prof. J.M. Nester for
useful comments.
The work of CMC was supported in part by the National Science Council
(Taiwan) under grant NSC 89-2112-M-008-016.

\end{multicols}
\end{document}